\documentclass[aps,prl,twocolumn,showpacs,10pt,floatfix,superscriptaddress,floatfix]{revtex4-1}
\usepackage{epsf}
\usepackage{amsfonts}
\usepackage{amssymb}
\usepackage{epsf}
\usepackage{enumerate,dsfont}
\usepackage{array}
\usepackage{makecell}
\usepackage{dcolumn,multirow}
\usepackage{latexsym}
\usepackage[utf8]{inputenc}
\usepackage{amsmath,mathtools}

\newcommand{\vk}{\vec k}

\newcommand{\ZZ}{\mathbb{Z}}

\newcommand{\be}{\begin{equation}}
\newcommand{\ee}{\end{equation}}

\renewcommand{\vec}[1]{\mathbf{#1}}

\begin{document}

\title{Reflection symmetric second-order topological insulators and superconductors}

\author{Josias Langbehn, Yang Peng, Luka Trifunovic, Felix von Oppen, and Piet W.
Brouwer}
\affiliation{Dahlem Center for Complex Quantum Systems and Physics Department, Freie Universit\"at Berlin, Arnimallee 14, 14195 Berlin, Germany}
\date{\today}

\begin{abstract}
Second-order topological insulators are crystalline insulators with a gapped bulk and gapped crystalline boundaries, but topologically protected gapless states at the intersection of two boundaries. Without further spatial symmetries, five of the ten Altland-Zirnbauer symmetry classes allow for the existence of such second-order topological insulators in two and three dimensions. We show that reflection symmetry can be employed to systematically generate examples of second-order topological insulators and superconductors, although the topologically protected states at corners (in two dimensions) or at crystal edges (in three dimensions) continue to exist if reflection symmetry is broken. A three-dimensional second-order topological insulator with broken time-reversal symmetry shows a Hall conductance quantized in units of $e^2/h$.
\end{abstract}
\maketitle

{\em Introduction.---}After the discovery of topological insulators and superconductors and their
classification for the ten Altland-Zirnbauer symmetry classes
\cite{hasan2010,bernevig2013,qi2011}, the concept of nontrivial topological
band structures has been extended to materials in which the crystal structure
is essential for the protection of topological phases \cite{chiu2016}. This
includes weak topological insulators \cite{fu2007}, which rely on the
discrete translation symmetry of the crystal lattice, and topological
crystalline insulators \cite{fu2011}, for which other crystal symmetries are
invoked to protect a topological phase. Whereas the original strong
topological insulators always have topologically protected boundary states,
weak topological insulators or topological crystalline insulators have
protected boundary states for selected surfaces/edges only.

In a recent publication, Schindler {\em et al.}\ \cite{schindler2017} proposed
another extension of the topological insulator (TI) family: a higher-order
topological insulator. Being crystalline insulators, these have well-defined
faces and well-defined edges or corners at the intersections between the faces.
An $n$th order topological insulator has topologically protected gapless states
at the intersection of $n$ crystal faces, but is gapped otherwise
\cite{schindler2017}. For example, a second-order topological insulator in two
dimensions ($d=2$) has zero-energy states at corners, but a gapped bulk and no
gapless edge states. Earlier examples of higher-order topological insulators
and superconductors {\em avant la lettre} appeared in works by Benalcazar
{\em et al.}
\cite{benalcazar2014,benalcazar2017,BenalcazarExtended} (see also
\cite{peng2017,PhysRevLett.110.046404}), who considered insulators and
superconductors with protected corner states in $d=2$ and $d=3$
\footnote{Reference \onlinecite{benalcazar2014} views the corner of a
three-dimensional lattice as a disclination of the surface. In contast, the
corner states of a two-dimensional second-order TI exist as the corners of a
``flat'' surface.}. Sitte {\em et al.} showed that a three-dimensional
topological insulator in a magnetic field of generic direction also acquires
the characteristics of a second-order topological Chern insulator, with chiral
states moving along the sample edges \cite{sitte2012}.

Since a second-order TI has a topologically trivial $d$-dimensional bulk, from
a topological point of view its boundaries are essentially stand-alone
$(d-1)$-dimensional insulators, so that topologically protected states at
corners (for $d=2$) or edges (for $d=3$) arise naturally as ``domain walls'' at
the intersection of two boundaries if these are in different topological
classes \cite{jackiw1976,hasan2010,hsieh2012}. Similarly, the classification of
$n$th order TIs derives from that of TIs in $d+1-n$ dimensions, {\em i.e.}, the
same classification of codimension $n$ topological defects \cite{teo2010} 
(see \cite{suppl} for a scattering-approach based classification of $n$th order TIs). Note
that, unlike for strong topological insulators and superconductors, which have
protected states at all boundaries, $n$th-order topological insulators and
superconductors have topologically protected states at the intersection of $n$
boundaries {\em only} if (some of) these boundaries are in different
topological classes; they do not necessarily have protected states at {\em all}
intersections of $n$ boundaries.

Apart from their role in stabilizing well-defined crystal faces, crystalline
symmetries are not required for the protection of higher-order TIs. However,
crystal symmetries can be a key to ensure that a natural surface termination
--- {\em i.e.}, a surface termination that respects the crystal symmetries ---
automatically leads to a nontrivial higher-order topological phase. For
example, Benalcazar {\em et al.} employed a combination of multiple reflection
symmetries \cite{benalcazar2017}, whereas Schindler {\em et al.} considered $C_4
{\cal T}$ symmetry, the product of a $\pi/2$ rotation and time reversal, 
as well as a model with reflection symmetry \cite{schindler2017}.

In this letter, we show that a single mirror symmetry is sufficient to
construct models for second-order topological insulators and superconductors in
$d=2$ and $d=3$ for all five Altland-Zirnbauer classes for which second-order
topological insulators are allowed. Reflection-symmetric topological
crystalline insulators were the first to be realized experimentally
\cite{hsieh2012,tanaka2012,xu2012}. A complete classification of
reflection-symmetric topological insulators and superconductors exists for all
ten Altland-Zirnbauer classes
\cite{chiu2013,morimoto2013,shiozaki2014,trifunovic2017}, and our construction
makes use of this classification.  Since the second-order topological phase
does not {\em require} the reflection symmetry for its existence --- see our
general remarks above and our detailed discussion below ---, in practice an
approximate reflection symmetry may well be sufficient, which significantly
enhances the prospects of an experimental realization.

\begin{figure}
\includegraphics[width=0.9\columnwidth]{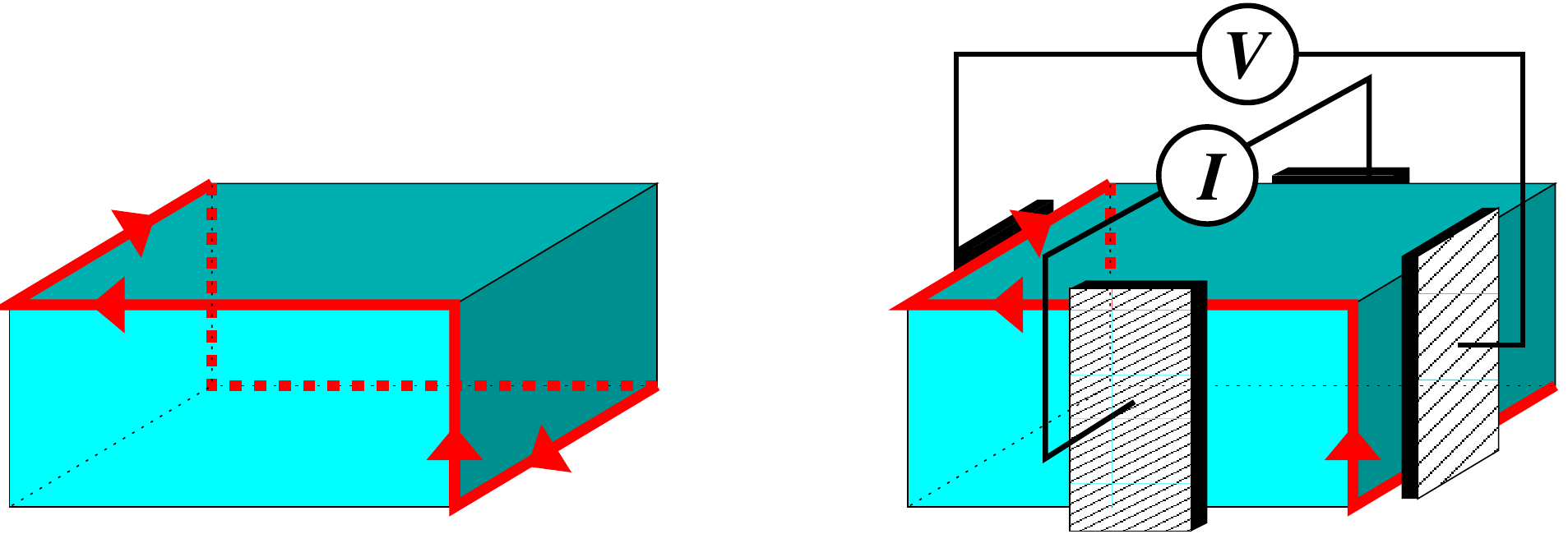}
\caption{\label{fig:1} A generic reflection-symmetric second-order topological insulator has chiral edge states winding around the crystal (left). If current and voltage contacts cover at most two neighboring edges, such a sample shows edge transport and, consequently, a quantized Hall effect in three dimensions (right).}
\end{figure}

An (approximately) reflection-symmetric three-dimensional second-order
topological insulator with broken time-reversal symmetry has chiral edge states
winding around the sample if none of its crystal faces is reflection symmetric,
see Fig.\ \ref{fig:1}(a). Despite being three-dimensional, such crystals show a
Hall effect with Hall {\em conductance} quantized in units of $e^2/h$, if
current and voltage probes are attached such that they touch a single sample
edge or two neighboring edges, see Fig.\ \ref{fig:1}(b) \cite{sitte2012}.
Such a quantized Hall
effect for a three-dimensional crystal is different from the
``three-dimensional quantized Hall effect'', which involves a topologically
nontrivial bulk state and a quantized Hall {\em conductivity}
\cite{halperin1987,montambaux1990,chalker1995,koshino2001,koshino2002}. Similarly, a
reflection- and time-reversal-symmetric three-dimensional second-order
topological insulator has a one-dimensional helical edge state winding around
the crystal, corresponding to a quantized spin Hall effect in three dimensions.
The experimental detection of such a quantized (spin) Hall effect should be an
unambiguous experimental signature of a second-order TI.

{\em Second-order topological insulators with reflecton symmetry.---}Since corners and edges follow the classification of one-dimensional and
two-dimensional topological insulators and superconductors, second-order
topological insulators with protected zero-energy corner states or with gapless
edge states can exist for selected Altland-Zirnbauer classes only, see Table
\ref{tab:1}. We now provide a systematic method to construct examples of
second-order topological insulators in all five nontrivial Altland-Zirnbauer
classes, using a single reflection symmetry ${\cal R}$ to ensure the presence
of topologically protected corner or edge states.

For each allowed Altland-Zirnbauer class, the construction requires (i) one or
more pairs of crystal faces that are mapped onto each other by reflection and
(ii) a reflection-symmetric topological crystalline phase which becomes trivial
if the reflection symmetry is broken. The nontrivial topology of the
corresponding Altland-Zirnbauer class in $d-1$ dimensions guarantees that the
reflection-symmetry-breaking mass term that gaps out any boundary states
existing in the presence of reflection symmetry is unique. Since this mass term
must be {\em odd} under reflection, the two surfaces listed under (i) must be
in different topological classes, ensuring the existence of zero energy
(gapless) states at at least two corners (edges). Table \ref{tab:1} lists the
reflection-symmetric phases that meet these criteria. We emphasize again that
the reflection symmetry is used to {\em construct} the second-order topological
insulator; it is itself not essential for the existence of zero-energy corner
states (for $d=2$) or gapless edge states (for $d=3$). The corner states (edge
states) are robust against a reflection-symmetry breaking perturbation, as long
as the bulk and edge (surface) gaps are not closed \footnote{In principle,
since the corners (edges) featured in the above construction may themselves be
reflection symmetric, the reflection symmetry may allow for additional states.
These, however, can be gapped out by any weak perturbation that breaks the
reflection symmetry.}.

Below we discuss three examples in detail: A two-dimensional second-order
topological superconductor with Majorana corner states (class D), a
three-dimensional second-order topological insulator with chiral edge states
(class A), and a three-dimensional second-order topological insulator with
helical edge states (class AII). In all cases we take reflection to map the
momentum component $k_1$ into $-k_1$, leaving the other momentum components
unchanged.

\begin{table}
  \begin{tabular}{|c||c|c|c||c|c||c|c|}
    \hline 
    Cartan & ${\cal T}$ & ${\cal P}$ & ${\cal C}$ & & $d=2$ & & $d=3$ \\\hline
    A & 0 & 0 & 0 & 0 & --- & $\ZZ$ & ${\cal R}$ \\\hline
    AIII & 0 & 0 & 1 & $\ZZ$ & ${\cal R}_+$ & 0 & --- \\\hline
    AI & 1 & 0 & 0 & 0 & --- & 0 & --- \\ \hline
    BDI & 1 & 1 & 1 & $\ZZ$ & ${\cal R}_{++}$ & 0 & --- \\ \hline
    D & 0 & 1 & 0 & $\ZZ_2$ & ${\cal R}_{+}$ & $\ZZ$ & ${\cal R}_{+} $\\ \hline
    DIII & -1 & 1 & 1 & $\ZZ_2$ & \begin{tabular}{@{}c@{}}${\cal R}_{++}$,\ ${\cal R}_{--}$ \\ ${\cal R}_{-+}$ \end{tabular} & $\ZZ_2$ & ${\cal R}_{++}$,\ ${\cal R}_{-+}$ \\ \hline
    AII & -1 & 0 & 0 & 0 & --- & $\ZZ_2$ & ${\cal R}_{+}$,\ ${\cal R}_{-}$ \\ \hline
    CII & -1 & -1 & 1 & $\ZZ$ & ${\cal R}_{++}$,\ ${\cal R}_{--}$\ & $0$ & --- \\ \hline
    C & 0 & -1 & 0 & 0 & --- & $\ZZ$ & ${\cal R}_+$,\ ${\cal R}_-$ \\ \hline
    CI & 1 & -1 & 1 & 0 & --- & 0 & --- \\ \hline
  \end{tabular}
\caption{\label{tab:1} The ten Altland-Zirnbauer classes are defined according to the presence or absence of time-reversal (${\cal T}$), particle-hole (${\cal P}$), and chiral symmetry (${\cal C}$). A nonzero entry indicates the square of the antiunitary symmetry operations ${\cal T}$ or ${\cal P}$. Two-dimensional and three-dimensional reflection-symmetric topological crystalline phases that can be used for the construction of second-order topological insulator/superconductor phases are listed in the right two columns, together with the corresponding topological classification. The symbols ${\cal R}_{\sigma_{\cal T}}$, ${\cal R}_{\sigma_{\cal P}}$, ${\cal R}_{\sigma_{\cal C}}$, and ${\cal R}_{\sigma_{\cal T},\sigma_{\cal P}}$ refer to a reflection operator that squares to one and commutes ($\sigma=+$) or anticommutes ($\sigma=-$) with ${\cal T}$, ${\cal P}$, or ${\cal C}$.}
\end{table}

{\em Second-order topological superconductor: class D.---}For a superconductor
with broken time-reversal and spin-rotation symmetry, particle-hole symmetry
${\cal P}$ is the only relevant symmetry operation. Without loss of generality
we may represent ${\cal P}$ by complex conjugation $K$ by working in a Majorana basis and the reflection
operation by the Pauli matrix $\sigma_1$ in an orbital subspace, so that the Hamiltonian $H(k_1,k_2)$
satisfies
\begin{align}
  H(k_1,k_2) &= -H^*(-k_1,-k_2) 
  \nonumber \\ &= \sigma_1 H(-k_1,k_2) \sigma_1.
\end{align}
Without reflection symmetry, class D in two dimensions has a $\ZZ$
classification, where the integer topological number counts the number of
chiral Majorana edge modes \cite{volovik1988,read2000}. Chiral Majorana modes
are incompatible with reflection symmetry. Instead, with reflection symmetry a
$\ZZ_2$ topological structure remains
\cite{chiu2013,morimoto2013,shiozaki2014}, counting the parity of the number of
helical ({\em i.e.}, counter-propagating) Majorana edge modes. A minimal
reflection-symmetric nontrivial gapless edge state at a reflection-symmetric
edge has edge Hamiltonian $H_{\rm edge} = v k_1 \sigma_3$, with $v$ the
velocity, which is a trivial edge in the absence of ${\cal R}$. Indeed, upon
breaking reflection symmetry, $H_{\rm edge}$ is gapped out by a unique mass
term $m \sigma_2$. An explicit model realizing this scenario is given by the
four-band tight-binding Hamiltonian,
\begin{align}
  H =& (M - \cos k_1 - \cos k_2) \tau_2 + \tau_1 \sigma_3 \sin k_1 + \tau_3 \sin k_2 \nonumber \\ &\, \mbox{}
  + \lambda \tau_2 \sigma_1,
  \label{eq:HD}
\end{align}
with $0 < |M| < 2$ and $\lambda$ numerically small. The physical implementaion
would require stacking two $p_x\pm i p_y$ superconductors~\cite{PLee} with
opposite chirality and coupling them in such a way to gap-out edge-modes at
non-reflection symmetric edges.

Computing the spectrum of low-lying excitations for a rectangular crystal with
edges at 45 degrees with respect to the symmetry axis we find a zero-energy
state well separated from higher-lying excitations by a gap. The wavefunction
of the zero-energy states is localized near the two sample corners where the
reflection-related edges meet, as shown in Fig.~\ref{fig:4}(a) and (b) for two
different arrangements of the reflection line with respect to the corner of the
crystal. The localized zero mode persists if the crystal is rotated, such that
there are no longer any reflection-related edges [Fig.~\ref{fig:4}(c)].

\begin{figure}
\includegraphics[width=0.9\columnwidth]{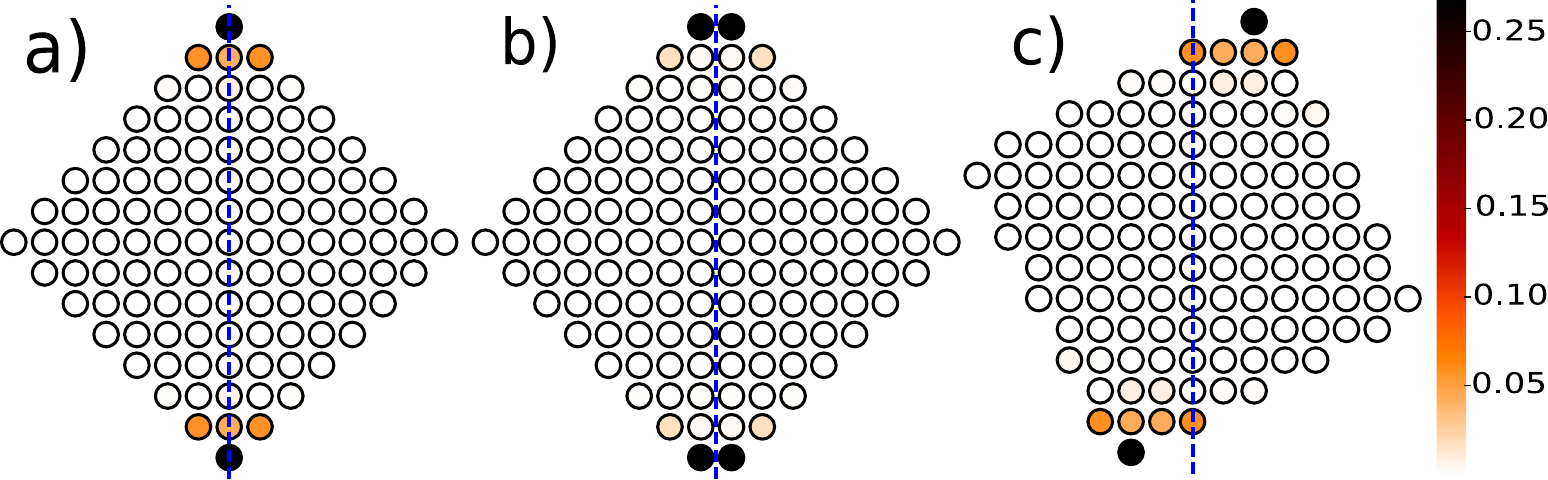}
\caption{\label{fig:4} Weight of the zero-energy wavefunction for the Hamiltonian (\ref{eq:HD}) with $M = 0.9$ and $\lambda = -0.25$ for three orientations of the crystal lattice. The reflection line is shown dashed.}
\end{figure}

{\em Second-order topological insulator in three dimensions: class A.---}In
three dimensions the presence of reflection symmetry allows for a topological
crystalline phase with an integer ``mirror Chern number'' enumerating gapless
surface states at reflection-symmetric surfaces
\cite{teo2008,chiu2013,morimoto2013,shiozaki2014}. Using $\sigma_1$ to
represent the reflection operation, such surface states have Hamiltonian
$H_{\rm surface} = v_1 k_1 \sigma_3 + v_2 k_2 \sigma_1$. The unique mass term
gapping out such surface states is $m \sigma_2$, which is odd under reflection.
Explicitly, one may consider the four-band Hamiltonian
\begin{align}
  H
  =&\, (M - \cos k_1 - \cos k_2 - \cos k_3) \tau_2 \sigma_1
  \label{eq:HA}
  \\ & \, \mbox{}
  + \sigma_3 \sin k_1 + \tau_1 \sigma_1 \sin k_2 
  + \tau_3 \sigma_1 \sin k_3
  + B \tau_2.\nonumber 
\end{align}
where $\sigma$ and $\tau$ are Pauli matrices in the space spanned by the
unit-cell orbitals two of which are even (odd) under reflection.  For $1 < M <
3$ and $B$ numerically small this Hamiltonian describes a three-dimensional
topologial insulator with a reflection-symmetric
time-reversal-symmetry-breaking term. To see how the bulk $B\tau_2$ term gives
rise to the $\sigma_2$ term at the non-reflection symmetric facet, we first note
that the details of the boundary do not have influence of the choice of the
$B$-term.  Thus to obtain facet-Hamiltonian for one of the $yz$-facets it is
sufficient to use low-energy expansion of the Hamiltonian (\ref{eq:HA}) and
model the boundary by the mass $M$ domain-wall along $x$-direction. This way we
immediately conclude that the projector onto the $yz$-facet Hamiltonian is
given by $\tau_2\sigma_2=\pm1$, therefore the term $B\tau_2$ acts like
$\sim\sigma_2$ within the surface subspace (the proportionality factor does
depends on the interface details).

\begin{figure}
\includegraphics[width=\columnwidth]{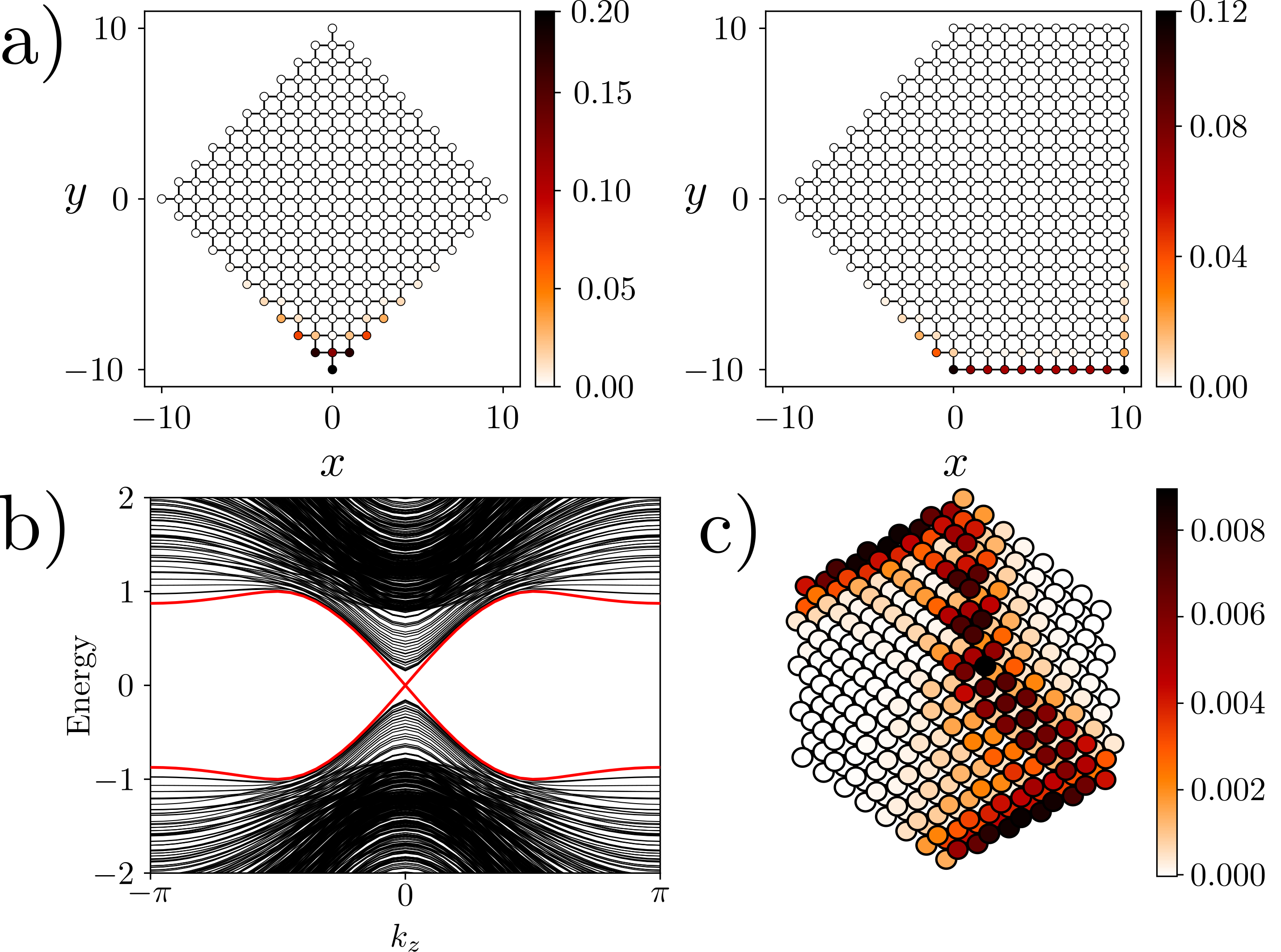}
\caption{\label{fig:5} (a) The crystal shape and the spatial profile of one of
the in-gap states. Left: when the two surfaces meet at the reflection
plane under a sharp angle, the in-gap state is well localized; right: one
surface is perpendicular to the reflection plane, the in-gap state becomes
completely delocalized on that surface. (b) Band structure of a rectangular
crystal with periodic boundary conditions in the $k_3$ direction, for the model
(\ref{eq:HA}) with parameters $M = 2$, $B = 0.2$.  (c) Weight of the zero-energy
chiral edge states for a finite lattice with generic orientation with respect to
the reflection
plane.}
\end{figure}

Figure \ref{fig:5}(b) shows the band structure of a rectangular crystal with
surfaces in the ($110$) and ($1\bar10$) direction, and periodic boundary
conditions in the $k_3$ direction. The two gapless chiral modes running in the
positive and negative $k_3$ direction are located at the intersection of the
($110$) and ($1\bar10$) surfaces related by ${\cal R}$, see
Fig.~\ref{fig:5}(a). We verified that the chiral edge states persist if the
crystal orientation is rotated by an angle less than 45 degrees and migrates to
the other pair of edges for larger rotation angles. Figure \ref{fig:5}(c) shows
the support of a chiral edge state for the above model Hamiltonian for a cubic
sample randomly oriented with respect to the reflection plane (so that none of
the facets are reflection symmetric). 

\begin{figure}
\includegraphics[width=0.95\columnwidth]{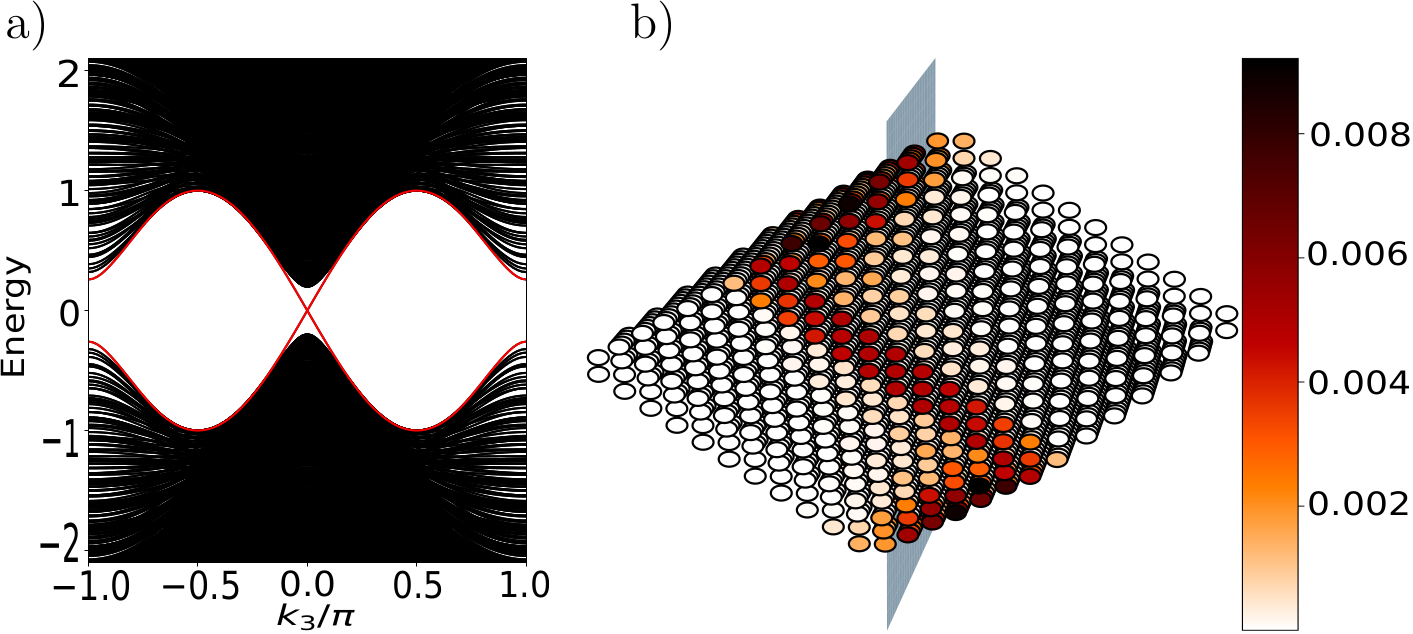}
\caption{\label{fig:6} (a) Band structure of a rectangular crystal with periodic boundary conditions in the $k_3$ direction, for the model (\ref{eq:HAII}) with parameters $M = 2$, $B = 0.8$. The crystal shape is the same as in Fig.\ \ref{fig:5}. (b) Weight of the zero-energy chiral edge states for a finite lattice with generic orientation with respect to the reflection plane.}
\end{figure}

{\em Second-order topological insulator in three dimensions: class AII.---}We
represent the time-reversal and reflection operations by $\sigma_2 K $ and
$\sigma_1$, respectively, so that $H(k_1,k_2,k_3)$ satisfies
\begin{align}
  H(k_1,k_2,k_3) =& \sigma_2 H(-k_1,-k_2,-k_3)^* \sigma_2 
  \nonumber \\ =& \mbox{}
  \sigma_1 H(-k_1,k_2,k_3) \sigma_1.
\end{align}
Hamiltonians with this symmetry have a $\ZZ$ topological classification
\cite{teo2008,chiu2013,morimoto2013,shiozaki2014} which enumerates the number
of surface states with Dirac-like dispersion $H_{\rm surface} = v_1 k_1
\sigma_3 + v_2 k_2 \sigma_1$ at reflection-symmetric surfaces. The unique
${\cal R}$-breaking mass term that gaps out such a pair of surface states
is $m \sigma_2 \tau_2$, where $\tau_2$ is an additional Pauli matrix. (A single
surface Dirac cone is protected by time-reversal symmetry.) Since this mass
term is odd under reflection, we expect an integer number of one-dimensional
helical states at the intersection of two surfaces related by ${\cal R}$. An
even number of helical states is unstable, however, to a local perturbation at
the edge and can be gapped out without closing the gaps in the sample bulk or
at the surfaces, consistent with the $\ZZ_2$ classification in Table
\ref{tab:1}. (At this point our classification differs from that of
Ref.~\cite{schindler2017}, which does not allow for reflection-symmetry
breaking perturbations at the crystal edge, thus arriving at a
$\ZZ$~\cite{trifunovic2017} classification.) As a specific example, we consider
the eight-band Hamiltonian
\begin{align}
  H
  =&\, [(M - \cos k_1 - \cos k_2 - \cos k_3) \tau_2 \sigma_1
  \nonumber \\ & \, \mbox{}
  + \sigma_3 \sin k_1 + \tau_1 \sigma_1 \sin k_2 
  + \tau_3 \sigma_1 \sin k_3] \rho_0
  \nonumber \\ & \, \mbox{}
  + B \tau_2 \rho_2,
  \label{eq:HAII}
\end{align}
where $\rho_{0,1,2}$ are an additional set of Pauli matrices, $1 < M < 3$, and
$B$ numerically small. One-dimensional band structure and weights for
zero-energy states are shown in Fig.\ \ref{fig:6} for the same geometry as in
the previous example.

{\em Conclusion.---}Although second-order topological insulators and
superconductors can exist without a topological crystalline bulk phase, the
existence of (approximate) crystalline symmetries can help in the construction
of models or in the identification of materials that realize these phases. 
For reflection-symmetric topological crystalline insulators, the published
literature has focused on reflection-symmetric surfaces, because these surfaces
harbor topologically protected surface
states~\cite{teo2008,hsieh2012,tanaka2012,xu2012,chiu2016}. We have shown that
there is a good reason to look at crystals with arbitrarily oriented surfaces,
because such crystals are good candidates for second-order TIs. Whereas the
surface states of reflection-symmetric topological crystalline insulators are
vulnerable to even weak perturbations that break the reflection symmetry, the
associated edge states are robust and persist as long as surface and bulk gaps
remain open. Combined with the unique prospect of isolated Majorana bound
states (for two-dimensional second-order topological superconductors) or
one-dimensional chiral modes and a quantized Hall effect (for three-dimensional
second-order TIs), higher-order TIs are a promising addition to the topological
materials family. Very recently, some early experimental realizations of
second-order topological insulators
appeared~\cite{Murani2017,2017arXiv170805015S,2017arXiv171003231P,2017arXiv170803647I},
where the Ref.~\cite{Murani2017} uses the bismuth nanowire which was previusly
shown to support edge states~\cite{deb2014}.

{\em Acknowledgments.---}This work was motivated by a colloquium on $C_4 {\cal
T}$-symmetric higher-order topological insulators by Titus Neupert at FU Berlin
and subsequent discussions. We also thank Andrei Bernevig and Max Geier for
discussions. We gratefully acknowledge support by projects A03 and C02 of the
CRC-TR 183 and by the priority programme SPP 1666 of the German Science
Foundation (DFG).

{\em Note added.---}After completion of this manuscript, we became aware of
related works by Ref. \cite{Song2017, Benalcazar2017b}.

\newpage

\begin{widetext}
\section*{Supplemental Material}

\section{Classification based on scattering theory} 

A variation of Laughlin's famous argument for a quantized Hall conductivity
\cite{laughlin1982}, can be used to classify corners, edges, etc.\ of $n$th
order topological insulators in $d$ dimensions in terms of their reflection
matrices. The classification closely follows the existing
reflection-matrix-based classification procedure for boundary states of
topological insulators in $d+1-n$ dimensions
\cite{simon2000,meidan2010,meidan2011b,fulga2011,fulga2012,fu2006}. We first
illustrate the reflection-matrix-based classification procedure for
second-order TIs in two dimensions, and then discuss the generalization to
higher orders and higher dimensions.

\begin{figure}[b]
\includegraphics[width=0.9\columnwidth]{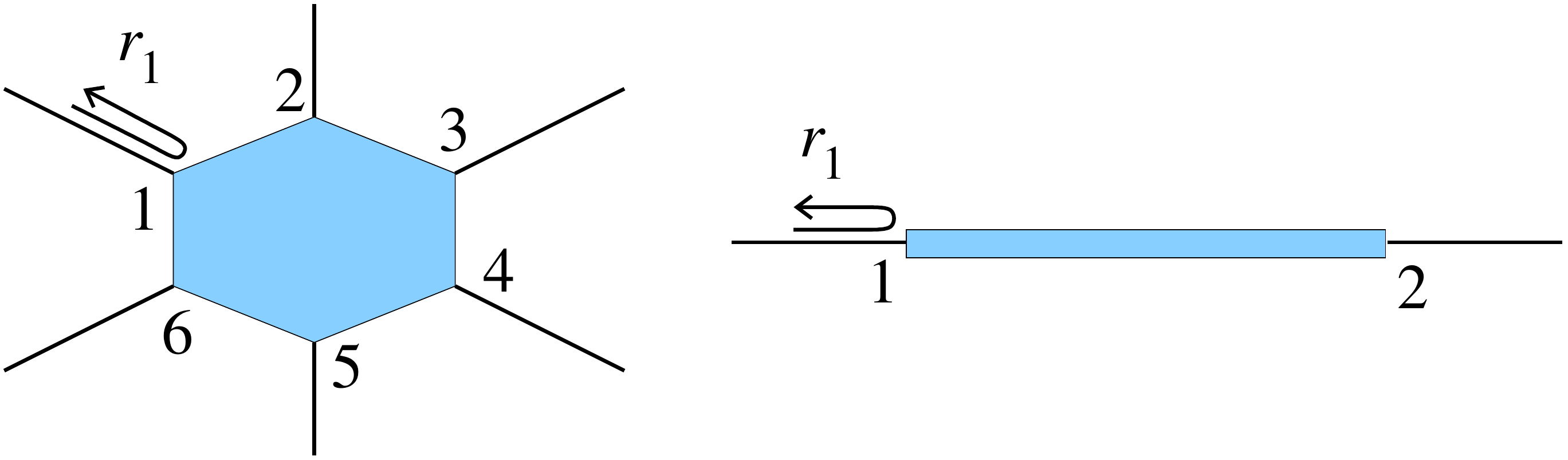}
\epsfxsize=0.95\hsize
\caption{\label{fig:2} Left: Two-dimensional second-order topological insulator (light blue), with one-dimensional leads attached to its corners. Right: One-dimensional topological insulator (light blue) with one-dimensional leads attached to its ends. In both cases each lead $j$, is described by a scattering matrix $r_j$.}
\end{figure}

For the classification of the corners of a two-dimensional crystal we attach a
one-dimensional lead to each corner, as shown in Fig.\ \ref{fig:2} (left). Each
lead has the number $g$ of degrees of freedom required for the corresponding
Altland-Zirnbauer class. (For example, $g = 2$ for a spinless Bogoliubov-de
Gennes Hamitonian, corresponding to particle and hole degrees of freedom.)
Since bulk and edges of the two-dimensional crystal are gapped, each lead is
described by a unitary reflection matrix $r_j$. There is a topologically
protected (zero-energy) corner state at the $j$th corner if and only if $r_j$
is ``topologically nontrivial'', where the precise meaning of ``topologically
nontrivial'' depends on the Altland-Zirnbauer class under consideration
\cite{fulga2011}. For example, for class D the topological number is
$\mbox{sign}\, \det r_j$ \cite{merz2002,akhmerov2011}. The same classification
procedure can be used for the two ends of a one-dimensional insulator, see
Fig.~\ref{fig:2} (right) \cite{fulga2011}, resulting in an identical
classification scheme. 

\begin{figure}[b]
\includegraphics[width=0.9\columnwidth]{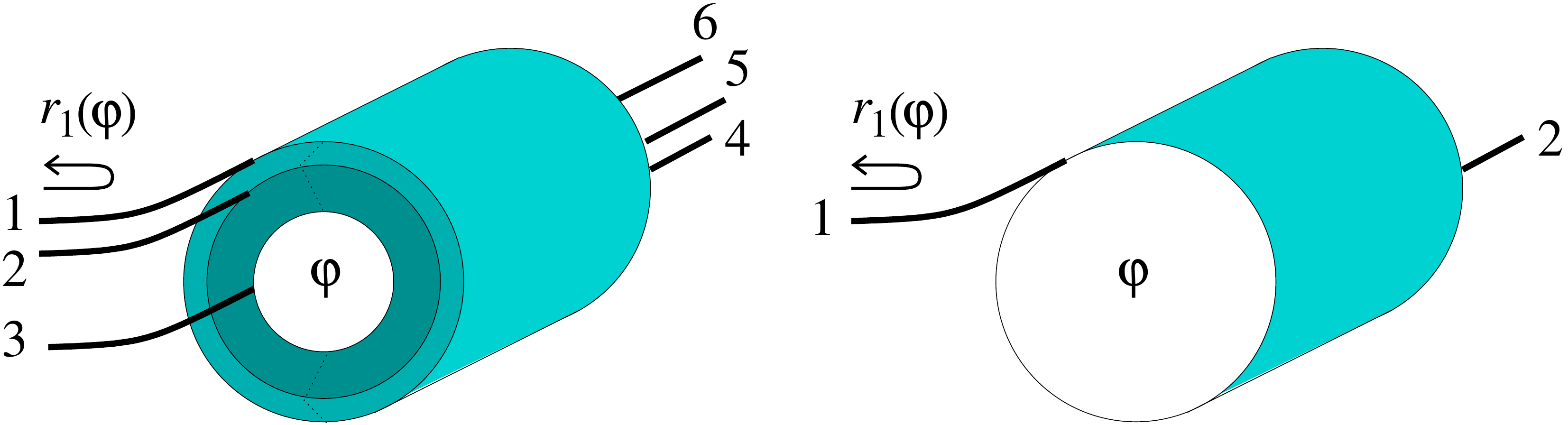}
\caption{\label{fig:3} Left: Three-dimensional second-order topological
insulator (light blue), with one-dimensional leads attached to its edges.
Right: Two-dimensional topological insulator (light blue) with one-dimensional
leads attached to its edge. In both cases periodic boundary conditions with
twist angle $\varphi$ are applied in the directin of the edge. Each each lead
$j$, is described by a unitary scattering matrix $r_j(\varphi)$.}
\end{figure}

The procedure is readily generalized to edges and corners of three-dimensional
topological insulators. For the classification of edges of three-dimensional
second-order topological insulators and superconductors, following Laughlin
\cite{laughlin1982}, we apply twisted periodic boundary conditions in the
direction along the crystal edge with twist phases $\varphi$, effectively
``rolling up'' the crystal into a cylinder of finite thickness, see Fig.\
\ref{fig:3} (left). We again attach one-dimensional leads at each of the
remaining sample edges, with the appropriate number of degrees of freedom $g$.
Again, each lead is described by a unitary reflection matrix $r_j$, which now
depends on the twist phase $\varphi$. Topologically protected gapless edge
modes exist at the $j$th crystal edge if and only if the one-dimensional family
of reflection matrices $r_j(\varphi)$ is topologically nontrivial, where, as
before, the precise definition of ``topologically nontrivial'' depends on the
Altland-Zirnbauer class \cite{simon2000,meidan2010,meidan2011b,fulga2012}. For
example, in class A, the winding number of $\det r(\varphi)$ for $0 < \varphi <
2 \pi$ determines the integer topological index \cite{simon2000}, whereas in
class AII the topological index is given by $\mbox{Pf}[r(\pi) \sigma_2]
\sqrt{\det r(0)}/\mbox{Pf}[r(0) \sigma_2] \sqrt{\det r(\pi)}$
\cite{meidan2011b,fulga2012}. Again, the same argument can be applied to, and
the same classification is obtained for the edge states of a standard
two-dimensional topological insulator, see Fig.~\ref{fig:3} (right).

We illustrate the procedure outlined above on a concrete example, the
three-dimensional Hamiltonian (3)  in class A in the main text. We consider a crystal
infinite in the $z$ direction, and with a cross section in the $xy$ plane as in
the inset of Fig.~\ref{fig:4}. We attach leads to all four vertical edges
(corresponding to the four corners in the cross section of Fig.\ \ref{fig:4})
and calculate the scattering matrix for electrons incident from these four
leads. Translation invariance in the $z$ direction allows us to fix the Bloch
wavenumber $k_z$. We verify that there is no transmission between different
leads, consistent with the fact that the bulk and the surfaces are gapped.
Figure~\ref{fig:4} shows that the total scattering phase $\phi_j = \mbox{arg}\,
\det r_j(k_z)$ for the two leads attached to the edges carrying chiral edge
state has non-zero winding number as a function of $k_z$, whereas there is no
winding number for the leads attached to corners without topologically
protected edge states.

\begin{figure}
\includegraphics[width=\columnwidth]{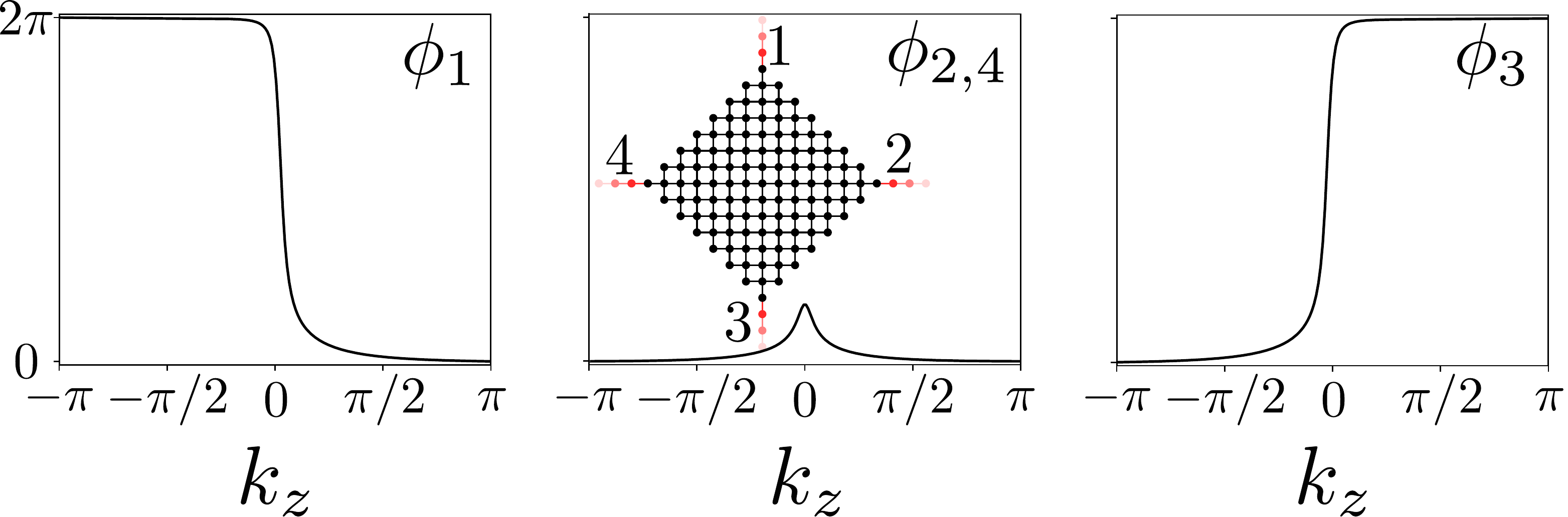}
\caption{\label{fig:4} The total scattering phase $\phi_j = \mbox{arg}\, \det
r_j(k_z)$ of the reflection matrices $r_j(k_z)$ for four leads $j=1,2,3,4$
attached to the four vertical edges of the three-dimensional
reflection-symmetric second-order topological insulator [Eq.~(3) in the main text]. The
inset shows a cross section in the $xy$ plane indicating the position of the
leads. Twisted periodic boundary conditions corresponding to the Bloch number
$k_z$ are applied in the $z$ direction. The system size is $30\times30$ sites
and the parameters take the values $m=2$, $B=0.2$. The scattering matrix is
calculated at the energy $\varepsilon=0.1$.}
\end{figure}

\section{\MakeLowercase{2d} second-order TI\MakeLowercase{s} without chiral and particle-hole symmetry}
Inspection of our classification Table~\ref{tab:longtable} shows that in 2d, the
non-trivial second-order TIs are possible only in the presence of either chiral
or particle-hole symmetry. This may seem to contradict the construction
presented in Ref.~\onlinecite{benalcazar2017}, where a quantized quadrupole moment
was predicted in the absence of these symmetries. 

The resolution of the above contradiction lies in the fact that we require the
edge-Hamiltonian (1d insulator) to be non-trivial TI (i.e. even in absence of
any crystalline symmetries). We can relax this requirement and only demand 
that the edge-Hamiltonian is non-trivial TCI, thus when the two reflection
symmetries are present, it is possible that a 1d edge-Hamiltonian is non-trivial
TCI even in the absence of chiral and particle-hole
symmetries~\cite{shiozaki2014,trifunovic2017}. We note that in this
scenario the existence of localized corner states in not guaranteed and one can
only guarantee existence of the corner charges~\cite{PhysRevB.94.165164}.
Another important difference is in the stability of the two above mentioned
types of higher-order TIs, the ones described in the Table~\ref{tab:longtable}
require only \textit{approximate} reflection symmetry as discussed in the main
text, whereas in example of Ref.~\cite{benalcazar2017} the quadrupole
quantization vanishes as soon as the reflection symmetry is broken.

\section{Discussion of the remaining symmetry classes}
Below we discuss the remaining symmetry classes for $d=2$ and $d=3$. For higher
dimensions ($d>3$) we use the isomorphisms between the corresponding
$K$-groups~\cite{trifunovic2017} to obtain which reflection symmetry needs to
be used, see Table~\ref{tab:longtable}.

\begin{table}
  \begin{tabular}{|c||c|c|c||c|c|c|c|c|c|c|c|}
    \hline 
    Cartan & ${\cal T}$ & ${\cal P}$ & ${\cal C}$ & $d=2$ & $d=3$ & $d=4$ & $d=5$ & $d=6$ & $d=7$ & $d=8$ & $d=9$\\\hline
    
    A & 0 & 0 & 0  & --- &  ${\cal R}(\ZZ)$  &--- & ${\cal R}(\ZZ)$ & --- & ${\cal R}(\ZZ)$ & --- & ${\cal R}(\ZZ)$ \\\hline
    
    AIII & 0 & 0 & 1  & ${\cal R}_+(\ZZ)$ &  ---  & ${\cal R}_{+}(\ZZ)$ & --- & ${\cal R}_{+}(\ZZ)$ & --- & ${\cal R}_{+}(\ZZ)$ & --- \\\hline
    \hline
    AI & 1 & 0 & 0  & --- &  ---  & --- & \begin{tabular}{@{}c@{}}${\cal R}_{+}$ \\ ${\cal R}_{-}$  \end{tabular}$(\ZZ)$& --- & \begin{tabular}{@{}c@{}}${\cal R}_{+}$ \\ ${\cal R}_{-}$  \end{tabular}$(\ZZ_2)$ & ${\cal R}_{+}(\ZZ_2)$ & ${\cal R}_{+}(\ZZ)$\\ \hline
    
    BDI & 1 & 1 & 1  & ${\cal R}_{++}(\ZZ)$ &  ---  & --- & --- & \begin{tabular}{@{}c@{}}${\cal R}_{++}$ \\ ${\cal R}_{--}$  \end{tabular}$(\ZZ)$ & --- & \begin{tabular}{@{}c@{}}${\cal R}_{++}$ \\ ${\cal R}_{--}$ \\ ${\cal R}_{+-}$ \end{tabular} $(\ZZ_2)$ & \begin{tabular}{@{}c@{}}${\cal R}_{++}$ \\ ${\cal R}_{+-}$  \end{tabular}$(\ZZ_2)$\\ \hline
    
    D & 0 & 1 & 0  & ${\cal R}_{+}(\ZZ_2)$ &  ${\cal R}_{+}(\ZZ) $  & --- & --- & --- & \begin{tabular}{@{}c@{}}${\cal R}_{+}$ \\ ${\cal R}_{-}$  \end{tabular}$(\ZZ)$ & --- & \begin{tabular}{@{}c@{}}${\cal R}_{+}$ \\ ${\cal R}_{-}$  \end{tabular}$(\ZZ_2)$\\ \hline
    
    DIII & -1 & 1 & 1  & \begin{tabular}{@{}c@{}}${\cal R}_{++}$ \\ ${\cal R}_{--}$ \\ ${\cal R}_{-+}$ \end{tabular}$(\ZZ_2)$  & \begin{tabular}{@{}c@{}}${\cal R}_{++}$ \\ ${\cal R}_{-+}$ \end{tabular} $(\ZZ_2)$ & ${\cal R}_{++}(\ZZ) $ & --- & --- & --- & \begin{tabular}{@{}c@{}}${\cal R}_{++}$ \\ ${\cal R}_{--}$  \end{tabular}$(\ZZ)$ & ---\\ \hline
    
    AII & -1 & 0 & 0 &  --- &  \begin{tabular}{@{}c@{}}${\cal R}_{+}$ \\ ${\cal R}_{-}$  \end{tabular}$ (\ZZ_2)$  & ${\cal R}_{+}(\ZZ_2) $ & ${\cal R}_{+}(\ZZ)$ & --- & --- & --- & \begin{tabular}{@{}c@{}}${\cal R}_{+}$ \\ ${\cal R}_{-}$  \end{tabular}$(\ZZ)$\\ \hline
    
    CII & -1 & -1 & 1  & \begin{tabular}{@{}c@{}}${\cal R}_{++}$ \\ ${\cal R}_{--}$ \end{tabular} $(\ZZ)$ & ---  & \begin{tabular}{@{}c@{}}${\cal R}_{++}$\\ ${\cal R}_{--}$ \\ ${\cal R}_{+-}$ \end{tabular}$(\ZZ_2)$ & \begin{tabular}{@{}c@{}}${\cal R}_{++}$ \\ ${\cal R}_{+-}$  \end{tabular}$(\ZZ_2)$ & ${\cal R}_{++}(\ZZ)$ & --- & --- & ---\\ \hline
    
    C & 0 & -1 & 0 & ---  & \begin{tabular}{@{}c@{}}${\cal R}_{+}$ \\ ${\cal R}_{-}$  \end{tabular} $(\ZZ)$ & --- & \begin{tabular}{@{}c@{}}${\cal R}_{+}$ \\ ${\cal R}_{-}$  \end{tabular}$(\ZZ_2)$& ${\cal R}_{+}(\ZZ_2)$ & ${\cal R}_{+}(\ZZ)$ & --- & ---\\ \hline
    
    CI & 1 & -1 & 1 &  ---  & --- & \begin{tabular}{@{}c@{}}${\cal R}_{++}$ \\ ${\cal R}_{--}$  \end{tabular}$(\ZZ) $ & --- & \begin{tabular}{@{}c@{}}${\cal R}_{++}$\\ ${\cal R}_{--}$ \\ ${\cal R}_{-+}$ \end{tabular}$(\ZZ_2)$ & \begin{tabular}{@{}c@{}}${\cal R}_{++}$ \\ ${\cal R}_{-+}$  \end{tabular}$(\ZZ_2)$ & ${\cal R}_{++}(\ZZ)$ & ---\\ \hline
  \end{tabular}
  \caption{\label{tab:longtable} Extension of Tab. I in the main text including higher dimensions.}
\end{table}

\subsection{$d=2$}
Below we discuss for each of the nontrival reflection-symmetric
Altland-Zirnbauer classes whether there are second-order topological insulators
if two of the faces are related by reflection symmetry. The ten
Altland-Zirnbauer classes are defined through the presence or absence of time
reversal symmetry ${\cal T}$, particle-hole symmetry ${\cal P}$, and chiral
symmetry ${\cal C}$, distinguishing the cases ${\cal T}^2 = \pm 1$ and ${\cal
P}^2 = \pm 1$. Explicitly, the three symmetry operations read
\begin{align}
  H(\vk) =&\, U_{\cal T} H(-\vk)^* U_{\cal T}^{\dagger}, \\
  H(\vk) =&\, - U_{\cal P} H(-\vk)^* U_{\cal P}^{\dagger}, \\
  H(\vk) =&\, - U_{\cal C} H(\vk) U_{\cal C}^{\dagger}, 
\end{align}
where $U_{\cal T}$, $U_{\cal P}$, and $U_{\cal C}$ are unitary matrices with
$U_{\cal T} U_{\cal T}^* = {\cal T}^2$ and $U_{\cal P} U_{\cal P}^* = {\cal
P}^2$. If ${\cal T}$ and ${\cal P}$ are both present, one has $U_{\cal C} =
U_{\cal T} U_{\cal P}^*$.

Reflection symmetry gives one more symmetry relation for $H$, 
\begin{equation}
  H(\vk) = U_{\cal R} H(R\vk) U_{\cal R}^{\dagger},
\end{equation}
where $R\vk = (-k_1,k_2)$ in two dimensions and $R\vk = (-k_1,k_2,k_3)$ in
three dimensions. The unitary matrix $U_{\cal R}$ is chosen such that the
reflection operation has the appropriate commutation or anticommutation
relations with the ${\cal T}$ and ${\cal P}$ operations. The precise choices
for the unitary matrices $U_{\cal T}$, $U_{\cal P}$, $U_{\cal C}$, and $U_{\cal
R}$ depend on the symmetry class under consideration and will be specified for
each symmetry class separately. Since the group structure follows from the
general classification procedure of second-order topological insulators and
superconductors, it is sufficient to show that the reflection-symmetric models
provide the generators.

\subsubsection{Class AIII}

{\em With reflection symmetry ${\cal R}_+$.---}Class AIII has chiral symmetry only, which we represent using $U_{\cal C} =
\sigma_3$. We represent the reflection symmetry using $U_{\cal R} = \sigma_3
\tau_3$. This class has a $\ZZ$ topological index, which counts the difference
between the number of helical edge states with positive and negative eigenvalue
of $\tau_3$. A ``minimal'' edge, which will serve as a generator for the
corresponding class of second-order TIs, has a single helical state with (say)
a positive eigenvalues of $\tau_3$, so that the reflection operation for edge
states is effectively represented by $U_{{\cal R},{\rm edge}} = \sigma_3$ and
the edge Hamiltonian is 
\begin{equation}
  H_{\rm edge} = v (\sigma_1 \cos \varphi + \sigma_2 \sin \varphi) k_1.
\end{equation}
In the absence of reflection symmetry, such an edge Hamiltonian has the unique
mass term $m (\sigma_1 \sin \varphi - \sigma_2 \cos \varphi)$, which is {\em
odd} under reflection. Hence, the intersection of two reflection-related edges
corresponds to a domain wall for the edge theory, which harbors a localized
state at zero energy.

\subsubsection{Class BDI}

{\em With reflection symmetry ${\cal R}_{++}$.---}Class BDI has time-reversal
symmetry and particle hole symmetry with ${\cal T}^2 = {\cal P}^2 = 1$. We here
represent time-reversal symmetry by complex conjugation $K$ and particle-hole
symmetry by $\tau_3 K$, so that the two-dimensional Hamiltonian satisfies
\begin{align}
  H(k_1,k_2) &= H(-k_1,-k_2)^* 
  \nonumber \\ &= -\tau_3 H(k_1,k_2) \tau_3.
\end{align}
A reflection symmetry commuting with ${\cal T}$ and ${\cal P}$ can be
represented by $U_{\cal R} = \tau_3 \sigma_3$, leading to the additional
constraint $H(k_1,k_2) = \tau_3 \sigma_3 H(-k_1,k_2) \sigma_3 \tau_3$. Class
BDI with ${\cal R}_{++}$ reflection symmetry has a $\ZZ$ classification. The
integer invariant counts the difference of the number of helical edge states
with positive and negative eigenvalues of $\tau_3$. The minimal
reflection-symmetric nontrivial edge has a single helical state with (say)
positive eigenvalue of $\tau_3$, so that effectively for the edge one
reflection is represented by $\sigma_3$. The corresponding minimal edge
Hamiltonian is $H_{\rm edge} = v k_1 \sigma_2$ and the unique mass term opening
a gap in $H_{\rm edge}$ is $m \sigma_1$, which is odd under reflection.

\subsubsection{Class D}

{\em With reflection symmetry ${\cal R}_{+}$.---}We refer to the discussion in the main text.

\subsubsection{Class DIII}

For Altland-Zirnbauer class DIII we choose the representations $U_{\cal T} =
\sigma_2$, $U_{\cal P} = 1$, so that
\begin{align}
  H(k_1,k_2) =&\, -H(-k_1,-k_2)^* \nonumber \\ =&\, 
  -\sigma_2 H(k_1,k_2) \sigma_2.
\end{align}
This symmetry class has a $\ZZ_2$ topological invariant in two dimensions,
which counts the parity of the number of helical Majorana edge modes. A single
helical Majorana edge mode has Hamiltonian
\begin{equation}
  H_1 = v k_1 \sigma_3 \label{eq:H1DIII}
\end{equation}
and can not be gapped out by any perturbation that preserves time-reversal and
particle-hole symmetry. A pair of Majorana modes, however, with edge
Hamiltonian
\begin{equation}
  H_2 = v k_1 \sigma_3 \tau_0,
\end{equation}
can be gapped out by the unique mass term $m \sigma_1 \tau_2$.  As in the case
of class AIII discussed above, the choice for the edge Hamiltonians $H_1$ and
$H_2$ is not unique, but different choices are related by an orthogonal
transformation, and the same orthogonal transformation needs to be applied to
the otherwise unique mass term.

{\em With reflection symmetry ${\cal R}_{++}$.---}We represent a reflection
operation that commutes with ${\cal T}$ and ${\cal P}$ by $U_{\cal R} = \tau_2
\sigma_2$. The corresponding topological classification is $\ZZ_2$. The
reflection symmetry is incompatible with the existence of only a single helical
Majorana mode. The mass term $m \sigma_1 \tau_2$ for a pair of Majorana modes
is odd under reflection operation, so that the generator of
DIII$^{{\cal R}_{++}}$ also serves as a generator for the corresponding class
of second-order topological superconductors. [Note that a quartet of helical
states, with edge Hamiltonian $H_4 = v k \sigma_3 \tau_0 \rho_0$, can be gapped
out, using the reflection-symmetric mass terms $m \sigma_1 \tau_1 \rho_2$ or $m
\sigma_1 \tau_3 \rho_2$, consistent with the $\ZZ_2$ index for this symmetry
class.]

{\em With reflection symmetry ${\cal R}_{--}$.---}We represent a reflection
operation that anticommutes with ${\cal T}$ and ${\cal P}$ by $U_{\cal R} =
\tau_3 \sigma_2$. With this reflection symmetry there is a $\ZZ$ topological
invariant, which counts the difference of the number of helical edge states
with positive and negative eigenvalues of $\tau_3$. For a ``minimal'' edge all
$\tau_3$-eigenvalues are (say) positive and one may effectively use the
representation $U_{\cal R} = \sigma_2$. Indeed, the combination of reflection
symmetry, particle-hole symmetry, and time-reversal symmetry then forbids any
mass term that would gap out helical edge modes. Only topological crystalline
phases with an even number of helical modes can be used for the construction of
a second-order topological insulator, since a single helical Majorana mode
corresponds to a strong topological phase. Each pair of helical modes leads to
a Majorana-Kramers pair at the intersection between reflection-related edges. A
pair of Majorana-Kramers pairs is unstable, however, to a local perturbation at
the sample corner and can be gapped out without closing the bulk and edge gaps. 

{\em With reflection symmetry ${\cal R}_{-+}$.---}In this case we take
$U_{\cal R} = \tau_3 \sigma_1$. The corresponding $\ZZ_2^2$ classification
counts the parity of the number of helical Majorana modes for each
$\tau_3$-eigenvalue. Topological phases with an odd number of helical Majorana
modes are topologically nontrivial already in the absence of reflection
symmetry and cannot be used for the construction of a second-order topological
superconductor. Since a pair of helical Majorana modes with the same
$\tau_3$-eigenvalue can be gapped out by the reflection-symmetric mass term
$\sigma_1 \rho_2$, the relevant topological crystalline phase has a pair of
helical Majorana modes with different $\tau_3$-eigenvalues: In this case the
edge modes are protected by reflection symmetry, since the reflection symmetry
forbids the unique mass term $m \sigma_1 \tau_2$.

\subsubsection{Class CII}
Altland-Zirnbauer class CII has ${\cal T}^2 = {\cal P}^2 = -1$, which is
implemented by choosing $U_{\cal T} = \sigma_2$ and ${\cal U}_{\rm P} =
\sigma_2 \tau_3$, so that $H(k_1,k_2)$ satisfies the symmetry constraints
\begin{align}
  H(k_1,k_2) &= \sigma_2 H(-k_1,-k_2)^* \sigma_2 
  \nonumber \\ &=
  - \tau_3 H(k_1,k_2) \tau_3.
\end{align}

{\em With reflection symmetry ${\cal R}_{++}$.---}We represent the reflection
symmetry by $U_{\cal R} = \tau_3 \rho_3$. The corresponding symmetry class has
a $2 \ZZ$ classification, see Refs.\
\cite{chiu2013,morimoto2013,shiozaki2014}, which counts difference of
``edge quartets'' with positive and negative eigenvalue of $\rho_3$. The
minimal nontrivial edge has a single quartet with positive eigenvalue of
$\rho_3$, so that effectively ${\cal R}$ is represented by $\tau_3$. The
corresponding edge Hamiltonian has the form
\begin{equation}
  H_{\rm edge} = i v k 
  \begin{pmatrix} 1 & 0 \\ 0 & n \end{pmatrix}
  \begin{pmatrix} 0 & i \\ -i & 0 \end{pmatrix}
  \begin{pmatrix} 1 & 0 \\ 0 & n^{\dagger} \end{pmatrix}
\end{equation}
where $n$ is a (real) quaternion of unit modulus, $n = n_0 + i \sum_{j=1}^{3}
n_j \sigma_j$, with $\sum_{j=0}^{3} n_j^2 = 1$. The unique mass term gapping
out $H_{\rm edge}$ is 
$$
  m \begin{pmatrix} 1 & 0 \\ 0 & n \end{pmatrix}
  \begin{pmatrix} 0 & 1 \\ 1 & 0 \end{pmatrix}
  \begin{pmatrix} 1 & 0 \\ 0 & n^{\dagger} \end{pmatrix},
$$%
and is odd under reflection. [Note that no mass terms can be added as long as
all edge quartets have the same $\rho_3$ eigenvalue, consistent with the $2
\ZZ$ topological index for this class.]

{\em With reflection symmetry ${\cal R}_{--}$.---}Proceeding as in the
previous case, we represent ${\cal R}$ by $\sigma_3$. For the helical edge we
choose the reflection-symmetric edge Hamiltonian

\begin{equation}
  H_{\rm edge} = v k_1 (\sigma_1 \cos \varphi + \sigma_2 \sin \varphi) \tau_1,
\end{equation}
and note that the unique mass term $m(\sigma_1 \sin \varphi - \sigma_2 \cos
\varphi) \tau_2$ is odd under reflection. (Note that upon doubling the number
of edge states, {\em e.g.}, taking $H_{\rm edge} = v k_1 \sigma_2 \tau_1
\rho_0$, a reflection-symmetric mass term $\tau_2 \rho_2$ exists, consistent
with the $\ZZ_2$ topological index for this
class \cite{morimoto2013,shiozaki2014}.)

\subsection{$d=3$}

We close with a discussion of the nontrival reflection-symmetric Altland-Zirnbauer classes in three dimensions, which become second-order topological insulators if two of the faces are related by reflection symmetry. 

\subsubsection{Class A}

{\em With reflection symmetry ${\cal R}$.---}We refer to the main text for a discussion.

\subsubsection{Class D}
{\em With reflection symmetry ${\cal R}_{+}$.---}As in the two dimensional
case, we represent particle-hole conjugation by complex conjugation and
reflection by $U_{\cal R} = \sigma_1$, so that $H(k_1,k_2,k_3)$ satisfies

\begin{align}
  H(k_1,k_2,k_3) &= -H(-k_1,-k_2,-k_3)^* 
  \nonumber \\ &= \sigma_1 H(-k_1,k_2,k_3) \sigma_1.
\end{align}
The corresponding symmetry class has a $\ZZ$ classification. A generator has
surface Hamiltonian
\begin{equation}
  H_{\rm surface} = v (k_1 \sigma_3 + k_2 \sigma_1),
\end{equation}
where we choose the (001) surface as a prototype for a reflection-symmetric
surface. The unique mass term gapping out these surface states is $m \sigma_2$,
which is odd under reflection. The intersection of two surfaces related by
reflection symmetry corresponds to a domain wall in the effective surface
theory, which hosts a chiral Majorana mode \cite{jackiw1976}.

\subsubsection{Class DIII}

We again choose the representations $U_{\cal T} = \sigma_2$, $U_{\cal P} = 1$, so that
\begin{align}
  H(k_1,k_2,k_3) =&\, -H(-k_1,-k_2,-k_3)^* \nonumber \\ =&\, 
  -\sigma_2 H(k_1,k_2,k_3) \sigma_2.
\end{align}
This symmetry class has a $\ZZ$ topological invariant in three dimensions,
which counts the number of Majorana Dirac cones at the surface. A
single Majorana Dirac cone has Hamiltonian
\begin{equation}
  H_2 = v (k_1 \sigma_1 + k_2 \sigma_3) \label{eq:H2DIII3d}
\end{equation}
and can not be gapped out by any perturbation that preserves time-reversal and
particle-hole symmetry. The same applies to multiple Dirac cones described by
the Hamiltonian (\ref{eq:H2DIII3d}), as long as they have the same chirality
(which is well defined because terms proportional to $\sigma_2$ are forbidden
by the chiral symmetry).

{\em With reflection symmetry ${\cal R}_{++}$.---}We represent the reflection
operation by $U_{\cal R} = \tau_1$. A single Majorana Dirac cone is
incompatible with reflection symmetry. A pair of Dirac cones with opposite
chirality is allowed under reflection, however, with surface Hamiltonian
\begin{equation}
  H_2 = v (k_1 \sigma_1 \tau_3 + k_2 \sigma_3 \tau_0).
  \label{eq:H2DIII3dOpp}
\end{equation}
Such a pair can be gapped out by the unique mass term $m \tau_2 \sigma_1$,
which is odd under reflection. [Note that upon doubling the number of Majorana
Dirac cones, setting $H_2 = v(k_1 \sigma_1 \tau_3 + k_2 \sigma_3 \tau_0)\rho_0$
the mass term $m = \rho_2 \sigma_1$ is allowed under reflection symmetry,
consistent with the $\ZZ_2$ topological classification for this symmetry
class.]

{\em With reflection symmetry ${\cal R}_{-+}$.---}This reflection symmetry can
be represented by the Pauli matrix $U_{\cal R} = \sigma_3$. This symmetry class
has a $\ZZ^2$ topological classification, where the integer invariants count
the number of Majorana Dirac cones of each chirality. As in the previous
example, a single pair of Majorana Dirac cones with opposite chirality can be
gapped out by the mass term $m \tau_2 \sigma_1$, which is odd under reflection.
[Note that, although this symmetry class allows for multiple Majorana Dirac
cones, an even number of helical Majorana modes at the intersection of two
reflection-related surfaces is unstable to local perturbations at the edge.]

{\em With reflection symmetry ${\cal R}_{+-}$.---}Although the reflection
symmetry ${\cal R}_{+-}$ allows for a nontrivial topological phase, these are
the same phases as in the absence of reflection symmetry, so that this symmetry
class can not be used to construct a second-order topological superconductor.
To see this explicitly, we choose the representation $U_{\cal R} = \tau_2
\sigma_3$. One then verifies that the reflection symmetry is compatible with an
integer number of pairs of Majorana Dirac cones of the same chirality, described by
the surface Hamiltonian
\begin{equation}
  H_{2 \pm} =  v (\pm k_1 \sigma_1 + k_2 \sigma_3) \tau_0.
\end{equation}
The $\ZZ$ topological invariant counts the difference of such Majorana Dirac
pairs of positive and negative parity. The nontrivial topological
realizations are also nontrivial without reflection symmetry, {\em i.e.},
they are a strong topological phase.

\subsection{Class AII}

{\em With reflection symmetry ${\cal R}_{-}$.---}See the discussion in the main text.

{\em With reflection symmetry ${\cal R}_{+}$.---}We choose the representation
$U_{\cal T} = \sigma_2$ and $U_{\cal R} = \sigma_1 \tau_2$, for which a pair of
surface Dirac cones, described by the Hamiltonian
\begin{equation}
  H_2 = v (k_1 \sigma_2 + k_2 \sigma_1) \tau_0,
\end{equation}
is gapped out by a unique mass term $m \sigma_3 \tau_2$, which is odd under
reflection. [A quartet of surface Dirac cones can be gapped out in the presence
of reflection symmetry, however, consistent with the $\ZZ_2$ topological index
for this symmetry class.]

\subsection{Class C}
This symmetry class has particle-hole symmetry squaring to $-1$, which is
represented by $\sigma_2 K$. Hamiltonians in this symmetry class then satisfy
the symmetry constraint 
\begin{equation}
  H(k_1,k_2,k_3) = - \sigma_2 H(-k_1,-k_2,-k_3)^* \sigma_2.
\end{equation}

{\em With reflection symmetry ${\cal R}_{+}$.---}We choose to represent
reflection by $U_{\cal R} = \tau_3$ and consider a pair of surface Andreev
Dirac cones with Hamiltonian
\begin{equation}
  H_2 = v (k_1 \tau_1 + k_2 \tau_3) \sigma_0
  \label{eq:H2C}
\end{equation}
This surface Hamiltonian is gapped out by the unique mass term $m \sigma_0
\tau_2$, which is odd under reflection. The intersection of two
reflection-related surfaces then has a domain wall in the effective surface
theory, which gives a chiral Andreev mode at the corresponding crystal edge.
[Note that multiple surface Majorana Dirac cones with Hamiltonian
(\ref{eq:H2C}) are protected in the presence of reflection symmetry, consistent
with the $2\ZZ$ topological index for this symmetry class.]

{\em With reflection symmetry ${\cal R}_{-}$---.}We represent the
anticommuting reflection operation by $U_{\cal R} = \sigma_1 \tau_3$. This
symmetry class has a $\ZZ_2$ topological index. The anticommuting reflection
symmetry is compatible with a pair of surface Andreev Dirac cones with
Hamiltonian (\ref{eq:H2C}), and the mass term $m \sigma_0 \tau_2$ remains odd
under reflection. Consistent with the $\ZZ_2$ topological index, a
quartet of Andreev Dirac cones with Hamiltonian $H_2 = v (k_1 \tau_1 + k_2
\tau_3) \sigma_0 \rho_0$ can be gapped out by a reflection-allowed mass term $m
\tau_2 \sigma_2 \rho_2$.

\end{widetext}

\end{document}